\documentclass[sn-mathphys]{sn-jnl}

\usepackage[utf8]{inputenc}

\usepackage{mathptmx} 
\usepackage{fancyhdr}
\usepackage[normalem]{ulem}
\usepackage[keeplastbox]{flushend}
\usepackage{subcaption}
\usepackage{pifont}
\usepackage{listings}
\usepackage{lstautogobble}
\usepackage{framed}
\usepackage{mdframed}

\definecolor{shadecolor}{gray}{0.9}

\definecolor{codegreen}{rgb}{0,0.6,0}
\definecolor{codegray}{rgb}{0.5,0.5,0.5}
\definecolor{codepurple}{rgb}{0.58,0,0.82}
\definecolor{backcolour}{rgb}{0.96,0.96,0.95}
\definecolor{magentacolour}{rgb}{0.85,0.10,0.50}
\definecolor{bluecolour}{rgb}{0.0,0.0,0.8}

\lstdefinestyle{snippet}{
    backgroundcolor=\color{white},
    commentstyle=\color{codegreen},
    keywordstyle=\bfseries,
    numberstyle=\tiny\color{codegray},
    stringstyle=\color{codepurple},
    basicstyle=\ttfamily\small,
    breakatwhitespace=true,
    breaklines=true,
    captionpos=t,
    keepspaces=true,
    numbers=none,
    numbersep=4pt,
    showspaces=false,
    showstringspaces=false,
    showtabs=false,
    tabsize=2
}

\lstdefinestyle{overview_comparison}{
    backgroundcolor=\color{white},
    commentstyle=\color{codegreen},
    keywordstyle=\bfseries,
    numberstyle=\tiny\color{codegray},
    stringstyle=\color{codepurple},
    basicstyle=\ttfamily\footnotesize,
    breakatwhitespace=true,
    breaklines=true,
    captionpos=t,
    keepspaces=true,
    numbers=none,
    numbersep=4pt,
    showspaces=false,
    showstringspaces=false,
    showtabs=false,
    tabsize=2,
    emph=[1]{
        Baseline, IRU
    },
    emphstyle=[1]{\bfseries},
}

\lstdefinestyle{api_code}{
    backgroundcolor=\color{backcolour},
    commentstyle=\color{codegreen},
    keywordstyle=\color{magentacolour},
    numberstyle=\tiny\color{black},
    stringstyle=\color{codepurple},
    basicstyle=\ttfamily\footnotesize,
    breakatwhitespace=true,
    breaklines=true,
    captionpos=t,
    keepspaces=true,
    numbers=left,
    numbersep=4pt,
    showspaces=false,
    showstringspaces=false,
    showtabs=false,
    tabsize=2,
    emph=[1]{
        addr_t, size_t,
        filter_op_t,
        uint32_t
    },
    emphstyle=[1]{\color{magentacolour}},
    emph=[2]{
        cudaMalloc, cudaFree,
        __global__, __shared__, __device__, __host__,
        __syncthreads
    },
    emphstyle=[2]{\color{codegreen}},
}

\lstdefinestyle{cuda_code}{
    backgroundcolor=\color{backcolour},
    commentstyle=\color{codegreen},
    keywordstyle=\color{magentacolour},
    numberstyle=\tiny\color{black},
    stringstyle=\color{codepurple},
    basicstyle=\ttfamily\small,
    breakatwhitespace=true,
    breaklines=true,
    captionpos=t,
    keepspaces=true,
    numbers=left,
    numbersep=4pt,
    showspaces=false,
    showstringspaces=false,
    showtabs=false,
    tabsize=2,
    emph=[1]{
        addr_t, size_t,
        filter_op_t,
        uint32_t
    },
    emphstyle=[1]{\color{magentacolour}},
    emph=[2]{
        cudaMalloc, cudaFree,
        __global__, __shared__, __device__, __host__,
        __syncthreads
    },
    emphstyle=[2]{\color{codegreen}},
    emph=[3]{
        load_iru
    },
    emphstyle=[3]{\color{bluecolour}},
}

\jyear{2022}%

\begin{document}

\title{Irregular Accesses Reorder Unit: Improving GPGPU Memory Coalescing for Graph-Based Workloads}


\author[1]{\fnm{Albert} \sur{Segura}}\email{asegura@ac.upc.edu}
\author[1]{\fnm{Jose Maria} \sur{Arnau}}\email{jarnau@ac.upc.edu}
\author[1]{\fnm{Antonio} \sur{Gonzalez}}\email{antonio@ac.upc.edu}

\affil[1]{\orgdiv{Departament d'Arquitectura de Computadors}, \orgname{Universitat Polit\`ecnica de Catalunya (UPC)}, \orgaddress{\street{Campus Nord, Jordi Girona 1-3}, \city{Barcelona}, \postcode{08034}, \country{Spain}}}




\maketitle

\section*{Abstract}

\begin{abstract}

GPGPU architectures have become the dominant platform for massively parallel workloads, delivering high performance and energy efficiency for popular applications such as machine learning, computer vision or self-driving cars. However, irregular applications, such as graph processing, fail to fully exploit GPGPU resources due to their divergent memory accesses that saturate the memory hierarchy.

To reduce the pressure on the memory subsystem for divergent memory-intensive applications, programmers must take into account SIMT execution model and memory coalescing in GPGPUs, devoting significant efforts in complex optimization techniques. Despite these efforts, we show that irregular graph processing still suffers from low GPGPU performance. 

We observe that in many irregular applications the mapping of data to threads can be safely changed. In other words, it is possible to relax the strict relationship between thread and data processed to reduce memory divergence. Based on this observation, we propose the Irregular accesses Reorder Unit (IRU), a novel hardware extension tightly integrated in the GPGPU pipeline. The IRU reorders data processed by the threads on irregular accesses to improve memory coalescing, i.e. it tries to assign data elements to threads as to produce coalesced accesses in SIMT groups. Furthermore, the IRU is capable of filtering and merging duplicated accesses, significantly reducing the workload. Programmers can easily utilize the IRU with a simple API, or let the compiler issue instructions from our extended ISA.

We evaluate our proposal for state-of-the-art graph-based algorithms and a wide selection of applications. Results show that the IRU achieves a memory coalescing improvement of 1.32x and a 46\% reduction in the overall traffic in the memory hierarchy, which results in 1.33x speedup and 13\% energy savings on average, while incurring in a small 5.6\% area overhead.

\end{abstract}

\section{Introduction}\label{s:introduction}

Since its popularization over the last decade, GPGPU architectures have enabled a broad domain of new applications by boosting linear algebra computations~\cite{bell2008efficient,li2011strassen}, empowering Big Data analytics~\cite{root2016mapd} and deploying Machine Learning~\cite{yan2020characterizing} in numerous fields such as speech recognition~\cite{chong2011efficient}, image processing~\cite{krizhevsky2012imagenet} and self-driving cars~\cite{kato2018autoware}. GPGPU architectures excel at processing highly-parallel throughput oriented applications, which exhibit regular execution and memory access patterns. However, applications that show irregular memory accesses or branch divergence suffer from severe underutilization of GPGPU's functional units~\cite{segura2019scu}. Graph processing algorithms are a popular example of irregular applications. Although graph-processing can potentially benefit from highly-parallel architectures, they process unstructured and irregular data, which results in sparse and unpredictable memory access patterns~\cite{nodehi2018tigr}. In addition, graph processing shows extremely low computation to memory access ratio~\cite{beamer2016understanding}, which further hinders GPGPU efficiency.

\begin{figure}[t!]
    \hspace{0.5cm}
    \begin{subfigure}[t]{0.395\linewidth}
        \centering
        \includegraphics[width=\linewidth]{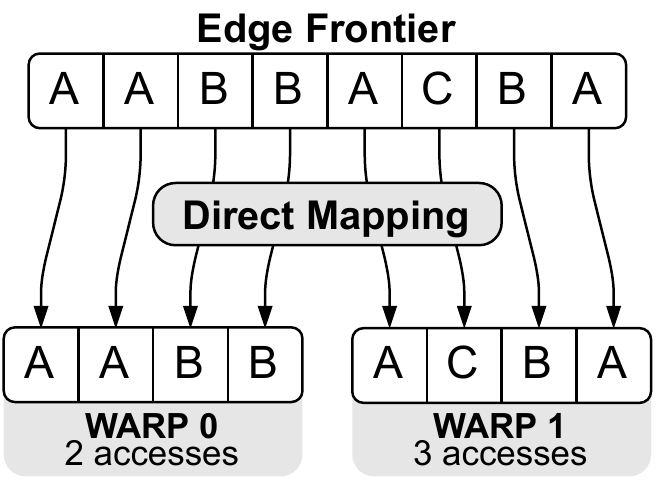}
        \subcaption{Baseline GPU}
        \label{f:frontier_baseline}
    \end{subfigure}%
    \hfill
    \begin{subfigure}[t]{0.395\linewidth}
        \centering
        \includegraphics[width=1\linewidth]{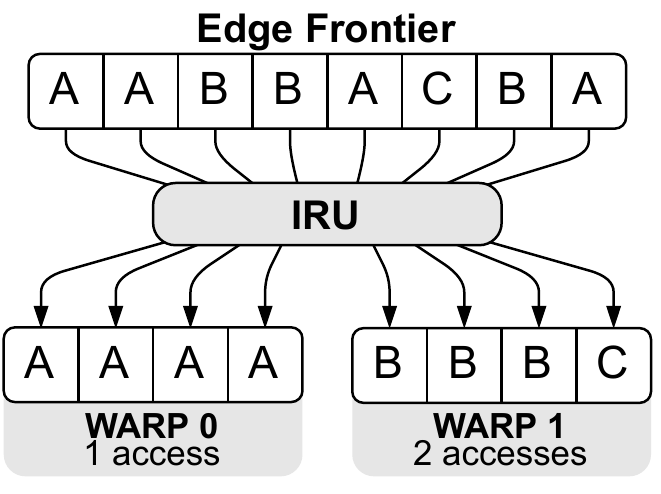}
        \subcaption{GPU with IRU}
        \label{f:frontier_iru}
    \end{subfigure}%
    \hspace{0.5cm}
\vspace{-0.05in}
\caption{Memory Coalescing improvement achieved by employing the IRU~(\ref{f:frontier_iru}) to reorder data elements that generate irregular accesses versus a Baseline GPU~(\ref{f:frontier_baseline}) execution. Assuming warp size of four threads for the sake of simplicity.}
\label{f:iru_coalescing_improvement}
\vspace{-0.1in}
\end{figure}

GPGPU programming models such as CUDA employ threads to exploit parallelism, each thread processing its own set of data while synchronizing with the rest to perform complex behaviors determined by the algorithm. The GPGPU pipeline handles the execution of warps, i.e. groups of threads in lock-step execution. The number of threads and the ability to coalesce the memory accesses within a warp are some of the key factors that determine the utilization of the GPU resources. The simplest way to exploit parallelism is to instantiate as many threads as data elements to process and directly assign each element to a given thread, as seen in Figure~\ref{f:frontier_baseline}. For a regular program, this assignment is highly effective at achieving high utilization of resources without inefficiencies (e.g. vector addition, where each thread processes consecutive data in memory achieving regular behavior). For programs exhibiting irregular memory accesses this simple assignment might cause utilization degradation, as the GPU is unable to achieve high memory coalescing in a warp, resulting in poor data locality (e.g. graph processing, where each thread processes a given node of the graph and has to fetch its adjacent ones).

To mitigate the aforementioned problems, GPGPU algorithms have to carefully consider the underlying hardware and adapt the algorithm to minimize branch divergence and improve memory coalescing, among other performance optimizations~\cite{burtscher2012quantitative,o2014microarchitectural}. Graph algorithms employ many such techniques, such as scan algorithms~\cite{sengupta2008efficient} which are leveraged for data compaction~\cite{billeter2009efficient} that gathers data to be accessed sparsely into a compacted data array, improving locality and memory coalescing. These techniques shift programmers effort from the algorithm to a hardware conscious programming requiring sound knowledge of the microarchitecture, significantly increasing development time while hampering code portability.

We claim that GPGPU programming models impose restrictions that hinder full resource utilization of irregular applications for several reasons. First, irregular programs such as graph processing algorithms consist of sparse and irregular memory accesses which have poor data locality and result in low memory coalescing, producing intra-warp memory divergence and significantly reducing GPU efficiency. Second, these issues are hard to improve without significant programmer effort to modify algorithms and data structures in order to better utilize the underlying hardware, which in some cases may not even be feasible and thus effectively limit the achievable performance. Ultimately, the programmer has to take into consideration ways to rearrange the data or change the mapping of data elements to threads to achieve better memory coalescing and higher GPU utilization, even if the relation of which threads process what data might not even be a restriction imposed by the algorithm, since the threads are primarily the means to expose parallelism.
Since GPGPU architectures and programming models are not designed to efficiently support sparse irregular programs, we propose to extend the GPU architecture to improve these workloads with a set of new instructions and their corresponding hardware support. We call this hardware the Irregular accesses Reorder Unit (IRU). The IRU is a small unit tightly integrated in the GPU, that is accessible through a set of new ISA instructions which can be used by the compiler or the programmer through a simple high-level API.

Our key idea is to relax the strict relation between a thread and the data that it processes. This allows the IRU to reorder the data serviced to the threads, i.e. to decide at run-time the mapping between threads and data elements to largely improve memory coalescing. Figure~\ref{f:iru_coalescing_improvement} shows conceptually how the IRU assigns data to the threads and achieves an improvement in memory coalescing against the baseline GPU. The IRU mapping improves the effectiveness of the memory coalescing hardware and the L1 data cache, as it results in better coalescing and locality, with subsequent improvements in the entire memory hierarchy, resulting in higher GPU utilization for irregular applications. In addition, the IRU performs simple preprocessing on the data (i.e filtering repeated elements), which reduces workload and allows for better utilization and further performance and energy improvements. In conclusion, the IRU optimizes irregular accesses requiring minimal support from programmers.

This paper focuses on improving the performance of irregular applications, such as graph processing, on GPGPU architectures. Its main contributions are the following:
\begin{itemize}
    \item We characterize the degree of memory coalescing and GPU utilization of modern graph-based applications. Our analysis shows that memory coalescing can be as high as 4 accesses per warp and GPU utilization as low as 13.5\%.
    \item We propose the IRU, a novel hardware unit integrated in the GPGPU architecture enabling improved performance of irregular accesses by reordering data serviced to each thread. We further extend the IRU to filter repeated elements, largely reducing GPU redundant workload for graph applications.
    \item We propose an ISA extension and API showing how modern graph-based applications can easily leverage the IRU.
    \item The GPU architecture with our IRU improves memory coalescing by a factor of 1.32x and reduces NoC traffic by 46\%, which result in 1.33x speedup and 13\% energy savings on average for a diverse set of graph-based applications. The IRU represents a small area overhead of 5.6\%.
\end{itemize}

The  remainder  of  this  paper  is  organized  as  follows. Section~\ref{s:background} reviews the challenges of irregular graph processing on GPGPU architectures. Section~\ref{s:hw} presents the architecture of the IRU and Section~\ref{s:api_unit} describes its API and usage for graph applications. Section~\ref{s:methodology} describes the methodology and Section~\ref{s:results} presents the evaluation. 
Section~\ref{s:related_work} reviews relevant related work and, finally, Section~\ref{s:conclusions} sums up the main conclusions.

\section{Irregular Applications on GPU Architectures}\label{s:background}

GPGPU architectures are tailored for compute intensive applications that feature regular execution and regular memory access patterns. 
GPU's high IPC is enabled by its Single-Instruction, Multiple-Threads (SIMT) pipeline, leveraging the advantage of decoding a single instruction for multiple threads, each operating on different data. The threads in a warp execute in a lock-step manner and, hence, to fully utilize the Execution Units (EU) applications must exhibit regular access patterns and control flow. Furthermore, to sustain high IPC, significant memory bandwidth is required which is accomplished with high Memory-Level Parallelism (MLP) leveraging warp-level coalescing and concurrent execution of many threads, increasing memory bandwidth at the expenses of increased latency.

On the other hand, for applications that show irregular behavior with unpredictable memory access patterns, GPGPU architectures are unable to provide enough memory bandwidth due to a huge portion of the threads generating uncoalesced accesses, which further hampers performance and results in low utilization of the EUs due to increased stalls.
In the worst case, a warp-level memory instruction requires 32 memory accesses (assuming warp size of 32 threads), as each thread may access a different cache line, whereas a perfectly coalesced warp-level memory instruction only requires one memory request (i.e. in case all the threads access the same cache line\footnote{If a sectored cache is used, perfect coalescing is only achieve if all the threads in a warp access the same sector. Note that if multiple sectors of a line are accessed, then multiple requests will be generated even if all the threads access the same cache line.}). Therefore, an irregular application may increase the requests to the memory subsystem by 32x compared to an application with regular access patterns and perfect memory coalescing. 


Not surprisingly, irregular applications increase the utilization of the LD/ST unit, the latency of memory instructions and the pressure on the L1 and the whole memory hierarchy. In addition, every warp instruction requires more resources to handle misses, such as miss status holding registers (MSHRs) and entries in the miss queue, a problem aggravated by GPUs small capacity ratio of cache lines per thread compared to CPUs. All these factors significantly increase the contention and conflict/capacity misses on the L1. Finally, the interconnection traffic also increases, L2 suffers from similar problems to the L1, and main memory accesses increase as a consequence of increased L2 misses.

Significant changes have to be applied to an algorithm and its data structures in order to reduce irregular accesses' overheads, and improve GPU efficiency. Generic approaches include the use of the shared memory in the Streaming Multiprocessors (SM) of the GPU, providing reduced latency and banked accesses of uncoalesced requests. Other approaches favor merging kernels, avoiding redundant memory requests at the cost of higher register usage. Graph algorithms use techniques such as data compaction, which reduce sparse accesses and improve locality by gathering sparse data in a compacted data array, as well as load balancing techniques that leverage collaborating threads which reduce branch and memory divergence.

Overall, irregular applications benefit from the high performance delivered by the massive parallelism of GPU architectures, but the architecture has significant bottlenecks that result in low performance for irregular algorithms. Significant programmer effort, code complexity and underlying hardware knowledge is required to create efficient GPU code for irregular applications such as graph processing algorithms.

\subsection{Graph Processing on GPGPU Architectures}

Many problems in Machine Learning~\cite{low2014graphlab,scarselli2008graph} and Data Analytics~\cite{zaharia2016apache} are modeled using graphs, which represent relationships between the elements on a set of data. GPGPU architectures enable fast parallel exploration and processing of the nodes and connections (i.e. edges) of a graph. Nonetheless, graph exploration is low-computation intensive~\cite{beamer2016understanding}, unstructured and irregular~\cite{lumsdaine2007challenges,xu2014graph} with sparse, irregular and highly unpredictable access patterns due to the irregular nature of the relationships expressed in a graph.

\lstset{style=snippet}
\begin{figure}[t!]
    \begin{subfigure}{0.5\linewidth}
        \centering
        \includegraphics[width=0.8\linewidth]{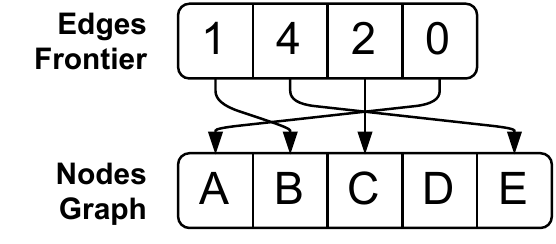}
    \end{subfigure}%
    \hfill
    \begin{subfigure}{0.490\linewidth}
        \centering
\begin{lstlisting}[language=Python]
for i in range(0, N):
  idx = edges_frontier[i] 
  out[i] = nodes_graph[idx]
\end{lstlisting}
        \vspace{0.05in}
    \end{subfigure}%
\vspace{0.05in}
\caption{The graph edges frontier produces irregular memory requests when accessing the nodes data in the graph. In a GPGPU, each thread may process one of the N elements in the edges frontier, i.e. perform one iteration of the loop. In this case, the access to \textit{edges\_frontier} shows high memory coalescing, as consecutive threads access consecutive memory locations. However, the access to \textit{nodes\_graph} array may result in high memory divergence depending on the indices.}
\label{f:irregular_access_example}
\end{figure}

A typical GPGPU graph processing algorithm starts in a given node and moves to adjacent nodes by traversing, or processing, that node edges. At this point, a new frontier (i.e. set of nodes or edges) is ready to be explored continuing this process iteratively until the whole connected graph is explored, or until the algorithm dictates it. Figure~\ref{f:irregular_access_example} shows how this process unfolds in a given iteration; each element of the edges frontier array (i.e. indices) points to the position to access in the nodes array to fetch for the next frontier data and continue the graph exploration. The pseudo-code shows the type of irregular access performed, which is an intrinsic part of graph exploration algorithms and a cause of the previously mentioned memory divergence. In this work we focus on common graph algorithms, in particular Breadth-First Search (BFS)~\cite{merrill2015high}, Single-Source Shortest Paths (SSSP)~\cite{davidson2014work} and PageRank (PR)~\cite{geil2014wtf}. 

GPGPU graph processing leverages many strategies to improve performance. First, data structures that efficiently represent the graph data in a compact manner using the Compressed Sparse Row (CSR)~\cite{bell2009implementing} format. Second, to cut down on sparse accesses, stream compaction algorithms~\cite{billeter2009efficient} are used to gather data in contiguous memory improving data locality and coalescing. Finally, 
load balancing techniques~\cite{merrill2015high} are used to leverage the threads in warps and thread blocks to cooperatively process data from the more processing demanding nodes, since the irregular graph connectivity of the nodes leads to nodes that largely differ in the number of edges. Although these techniques improve GPGPU efficiency for graph processing, significant changes are required to implement these optimizations and reduce the GPGPU architecture bottlenecks for irregular applications. Despite all these efforts, we observe that modern graph applications experience significantly low memory coalescing of 4 accesses per warp, leading to a low 13.5\% utilization of GPU resources. In the next section we present a novel hardware unit that improves memory coalescing and GPU performance for irregular workloads, while requiring minimal changes in the applications.

\section{Irregular accesses Reorder Unit}\label{s:hw}

In this section, we introduce the Irregular accesses Reorder Unit (IRU), which improves performance of irregular workloads such as graph applications on GPGPU architectures. Low GPGPU performance for graph processing is mainly due to the uncoalesced memory accesses, that result in large memory traffic and put significant pressure on the memory hierarchy. 
Our proposal improves GPGPU performance for graph-processing by assigning data elements (nodes/edges) that produce coalesced memory accesses to the threads in the same warp. This dynamic reordering of elements is done in hardware and it can be easily used by programmers, that only have to indicate when it is safe to change the mapping of data elements to threads.



\begin{figure}[t!]
    \hspace{-0.25in}
    \centering
    \includegraphics[width=0.75\linewidth]{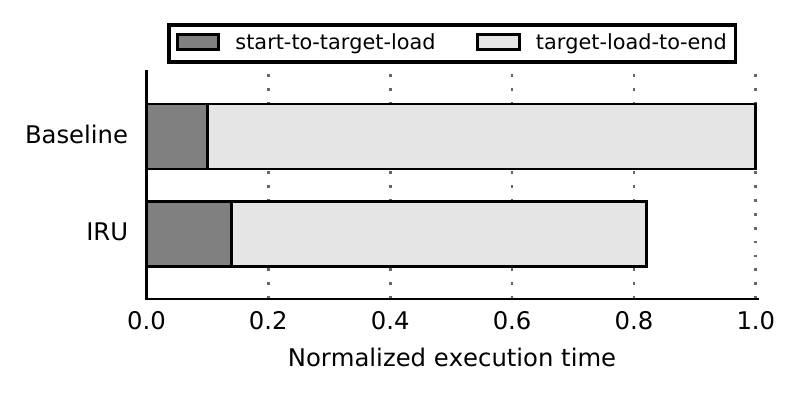}
    \null
    \vspace{0.125in}
    \caption{Warp average normalized execution with and without IRU. The dark bar indicates execution time until the target load is serviced, and the light bar from service to finalization. Processing a load instruction with the IRU is slower as it has to reorder data elements to identify indices that target the same cache line. However, once the indices are sent to the SMs, the remaining execution is faster as subsequent memory accesses used the IRU-prepared indices that result in higher memory coalescing.}
    \label{f:iru_gains}
\end{figure}

\begin{figure*}[t!]
    \begin{subfigure}[h]{0.22\linewidth}
        \centering
        \vspace{0.2in}
        \begin{tikzpicture}
        \node[anchor=south west,inner sep=0] at (0,0) {
        \includegraphics[width=0.9\linewidth]{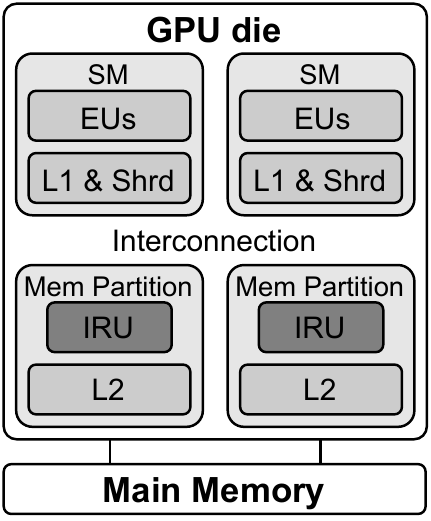}};
        \end{tikzpicture}
        \subcaption{GPU architecture with IRU.}
        \label{f:gpu_overview}
    \end{subfigure}%
    \hfill
    \begin{subfigure}[h]{0.36\linewidth}
        \centering
        \vspace{0.1in}
        
        \begin{tikzpicture}
        \node[anchor=south west,inner sep=0] at (0,0) {\includegraphics[width=0.9\linewidth]{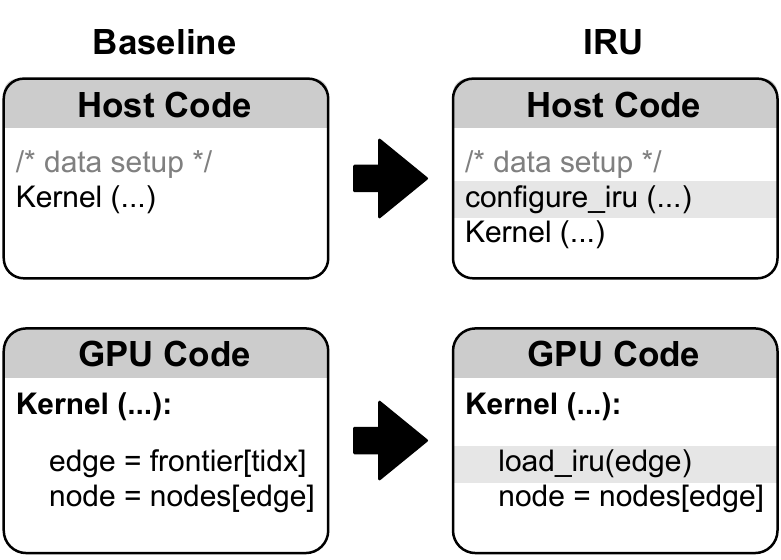}};
        \node at (0.03\linewidth,0.12\linewidth) {\ding{192}};
        \node at (0.03\linewidth,0.07\linewidth) {\ding{194}};
        \node at (0.547\linewidth,0.12\linewidth) {\ding{193}};
        \node at (0.547\linewidth,0.07\linewidth) {\ding{195}};
        \end{tikzpicture}
        \subcaption{Irregular access code modifications.}
        \label{f:iru_integration_code_comparison}
    \end{subfigure}%
    \hfill
    \begin{subfigure}[h]{0.36\linewidth}
        \centering
        \vspace{0.1in}
        \null\hfill
        \begin{tikzpicture}
        \node[anchor=south west,inner sep=0] at (0,0)
        {\includegraphics[width=0.9\linewidth]{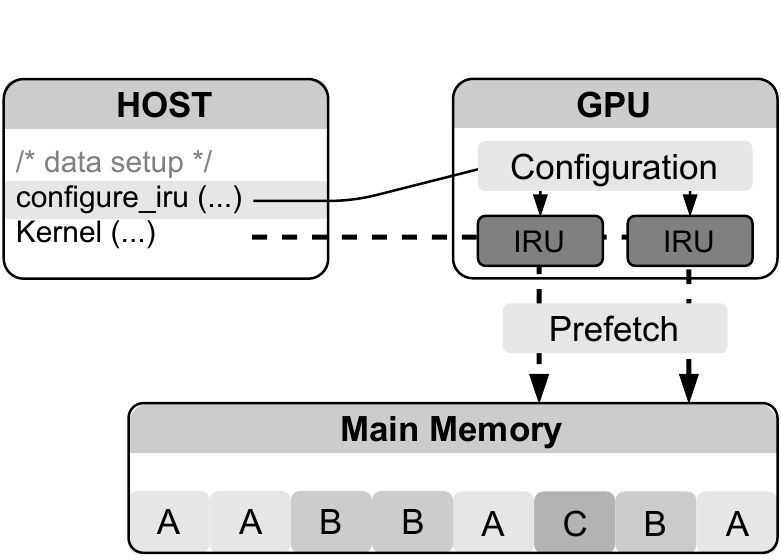}};
        \end{tikzpicture}
        \hfill\null
        \subcaption{IRU Configuration and Initialization.}
        \label{f:iru_integration_config}
    \end{subfigure}%
    
    \begin{subfigure}[b]{0.26\linewidth}
    
        \null\hfill
        \begin{tikzpicture}
        \node[anchor=south west,inner sep=0] at (0,0) {
        \includegraphics[width=0.92\linewidth]{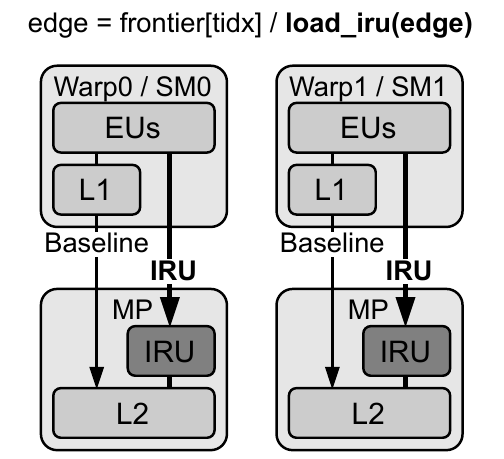}};
        \node at (0.00\linewidth,2.5) {\ding{192}};
        \node at (0.92\linewidth,2.5) {\ding{193}};
        \end{tikzpicture}
        \hfill\null
        \vspace{-0.15in}
        \subcaption{Irregular accesses indices retrieval.}
        \label{f:iru_integration_exec_indexes}
    \end{subfigure}%
    \hfill
    \begin{subfigure}[b]{0.72\linewidth}
        
        \hfill
        \begin{tikzpicture}
        \node[anchor=south west,inner sep=0] at (0,0) {
        \includegraphics[width=0.9\linewidth]{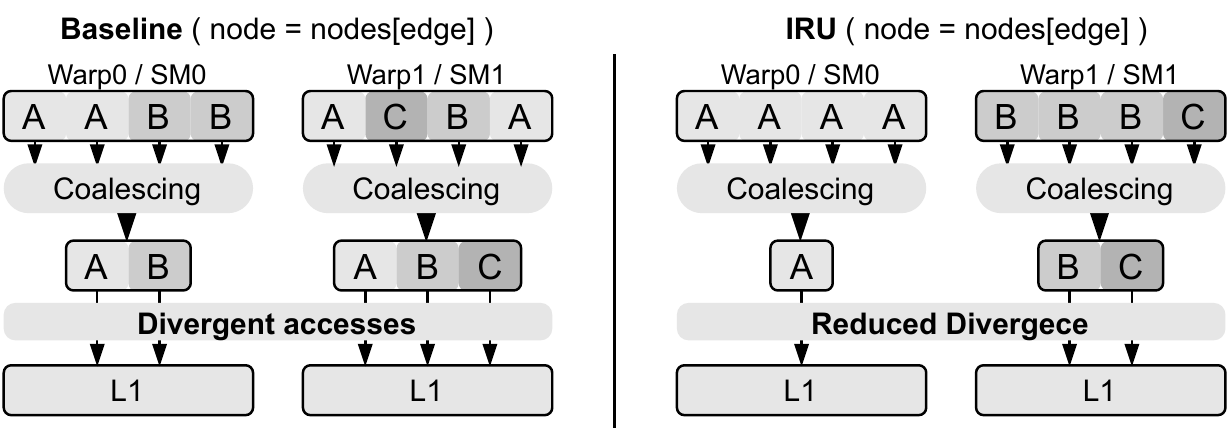}};
        \node at (0.02\linewidth,2.6) {\ding{194}};
        \node at (0.55\linewidth,2.6) {\ding{195}};
        \end{tikzpicture}
        \hfill\null
        \subcaption{Improvement of irregular accesses executed with the IRU reordered indices compared to the Baseline.}
        \label{f:iru_integration_exec_divergence}

    \end{subfigure}%
\caption{IRU integration with the GPU at different levels: architectural (a), program model (b) and execution (c,d,e). The execution shows how the program (b) works on the Baseline and the IRU-enhanced GPU with two warps and data from Figure~\ref{f:iru_coalescing_improvement}.}
\label{f:iru_overall_integration}
\end{figure*}
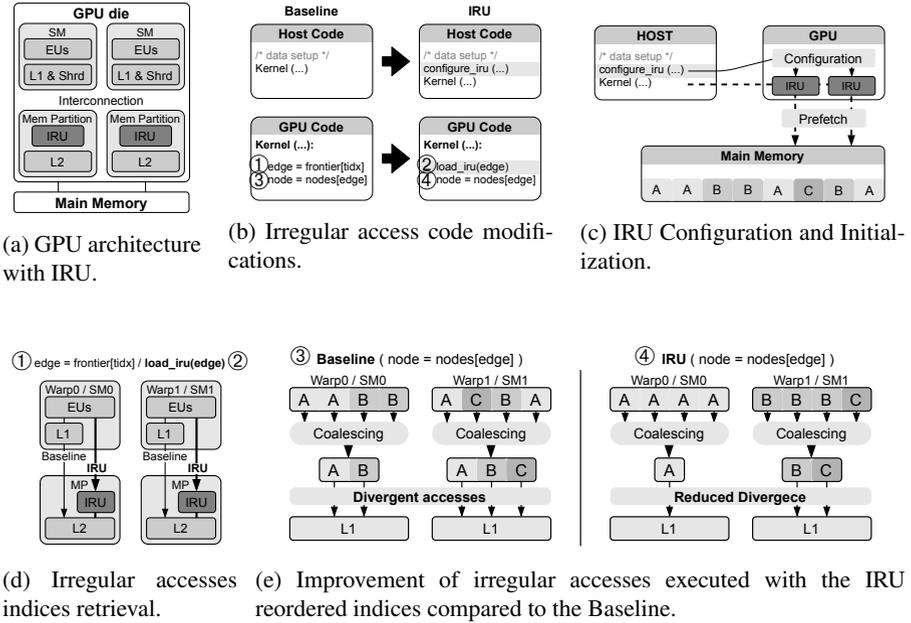

In this paper, we propose to extend the GPGPU with the IRU to reduce the overheads caused by irregular accesses. The IRU is a compact and efficient hardware unit integrated into the Memory Partition (MP) of the GPU architecture as shown in Figure~\ref{f:gpu_overview}, which incurs in very small energy and area overheads. The IRU leverages the observation that GPU programs employ threads to convey parallelism; being in many cases independent of the data that they process. The main goal of the IRU is to process, reorder and redistribute the indices used to perform irregular memory accesses. The reordering collocates indices that access the same memory block and services them to a requesting warp, reducing the memory divergence of irregular accesses. In turn, the improved memory coalescing reduces congestion of the resources of the LD/ST unit, L1, interconnection, L2 and main memory, significantly reducing the pressure on the memory subsystem. In addition, the reordering is performed across all the indices accessed by all the SMs, and hence, collocating irregular accesses potentially gathers data obtained by irregular accesses in a single or fewer SMs, thus further reducing interconnection traffic and L1 data thrashing.

Figure~\ref{f:iru_gains} shows the average normalized execution of a warp in a baseline GPU against one with the IRU. The dark bar indicates the execution time until the load processed and reordered by the IRU is serviced, while the light bar shows the normalized time until finalization. As it can be seen, processing a load instruction through the IRU is slower as it has to reorder the data elements to identify indices that target the same cache line. Furthermore, it also identifies and filters duplicated elements. For this reason, the \textit{start-to-target-load} time in Figure~\ref{f:iru_gains} is larger with the IRU than in the baseline GPU. However, once the indices are reordered and sent to the warps, the remaining execution is significantly faster as memory coalescing is largely improved for the subsequent memory accesses.
Therefore, the low overhead incurred by the IRU servicing the load is effectively offset by the performance gain achieved from the reduction of the overheads due to the memory divergence.

The IRU processes the indices of a target irregular instruction, with the objective to improve its coalescing. Additionally, the elements processed can contain more data than just the indices, as mandated by the API described in Section~\ref{s:api_unit}. While these data are not used for the IRU coalescing logic, it is responsible to fetch and send these additional data to the SM.

\lstset{style=overview_comparison}

\subsection{GPU Integration}

The IRU integration into the GPU is covered in Figure~{\ref{f:iru_overall_integration}}, showing architectural~{\ref{f:gpu_overview}}, programming~{\ref{f:iru_integration_code_comparison}} and execution~{\ref{f:iru_integration_config}}~-{\ref{f:iru_integration_exec_divergence}} integration. The execution shows how the Baseline and the IRU modified GPU programs in Figure~{\ref{f:iru_integration_code_comparison}}, operate with the two warps and data from Figure~{\ref{f:iru_coalescing_improvement}}. 
The Baseline program performs a regular access~\ding{192} to gather indices that are then used for an irregular access. The IRU modified code performs the same operation but using the IRU hardware with the \textit{load\_iru} operation~\ding{193}, which is part of the IRU API presented in Section~{\ref{s:api_unit}}.
The baseline code is executed by the GPU as follows. First, the two warps retrieve the indices performing regular accesses to the L1, i.e. consecutive threads in a warp access consecutive memory addresses. Afterwards, Figure~{\ref{f:iru_integration_exec_divergence}} shows how they perform irregular accesses to the L1 with the retrieved indices which, due to the high divergence, result in many memory requests~\ding{194}.

In contrast, the IRU program first introduces a configuration step performed on the host, shown in Figure~{\ref{f:iru_integration_config}}, that provides data of the irregular accesses to optimize. The configuration required for this program consists of the base address and data type of the irregular accessed data, and the indices array and total number of irregular accesses. Further IRU capabilities are enabled and used with optional parameters employed on overloaded functions, reviewed in Section~{\ref{s:api_unit}}. Next, when the kernel execution starts, the IRU triggers the prefetching of the indices from L2 and memory, which are then automatically reordered in the IRU hash. The IRU activity is overlapped with the execution of the kernel, and disabled when all the data is processed. Furthermore, the IRU is disabled for kernels that do not require reordering indices.

Regular execution proceeds until encountering the \textit{load\_iru} operation, at which point the warps retrieve the indices performing requests directly to the IRU, bypassing the L1 as seen in Figure~{\ref{f:iru_integration_exec_indexes}}. The IRU replies with reordered indices either instantly, if 32 indices that target the same cache line are available, or otherwise after a timeout to avoid starvation. In case the timeout is triggered, the 32 indices will not be collocated to the same cache line, but the IRU will do the best attempt to provide indices that result in the lowest memory divergence with the available elements. Finally, the warps perform the irregular access that was the target of the optimization~\ding{195}. This access is performed with the IRU reordered indices which achieves reduced divergence, performing less accesses than the baseline program, as depicted in Figure~{\ref{f:iru_integration_exec_divergence}}. Note that we assume a warp size of 4 threads in Figure~{\ref{f:iru_integration_exec_divergence}} for the sake of simplicity, but we use warp size of 32 for our experimental evaluation.

The IRU is first configured by the programmer with a host function, described in Section~\ref{s:api_unit}, that provides the IRU information about the data that has to be reordered. Once configured, the kernel is launched and the IRU begins its operation prefetching the required data and reordering the elements according to the configuration, with the aim of maximizing memory coalescing. IRU operation is autonomous and no further programmer intervention is required to reorder the elements. The SM kernel retrieves the reordered indices used by the irregular accesses using the API described in Section~\ref{s:api_unit}, requiring minimal changes to the code.

\subsection{Hardware Overview and Processing}
\begin{figure*}[t!]
    \centering
    \begin{subfigure}[h]{0.385\linewidth}
        \centering
        \begin{tikzpicture}
        \node[anchor=south west,inner sep=0] at (0,0) {\includegraphics[width=\linewidth]{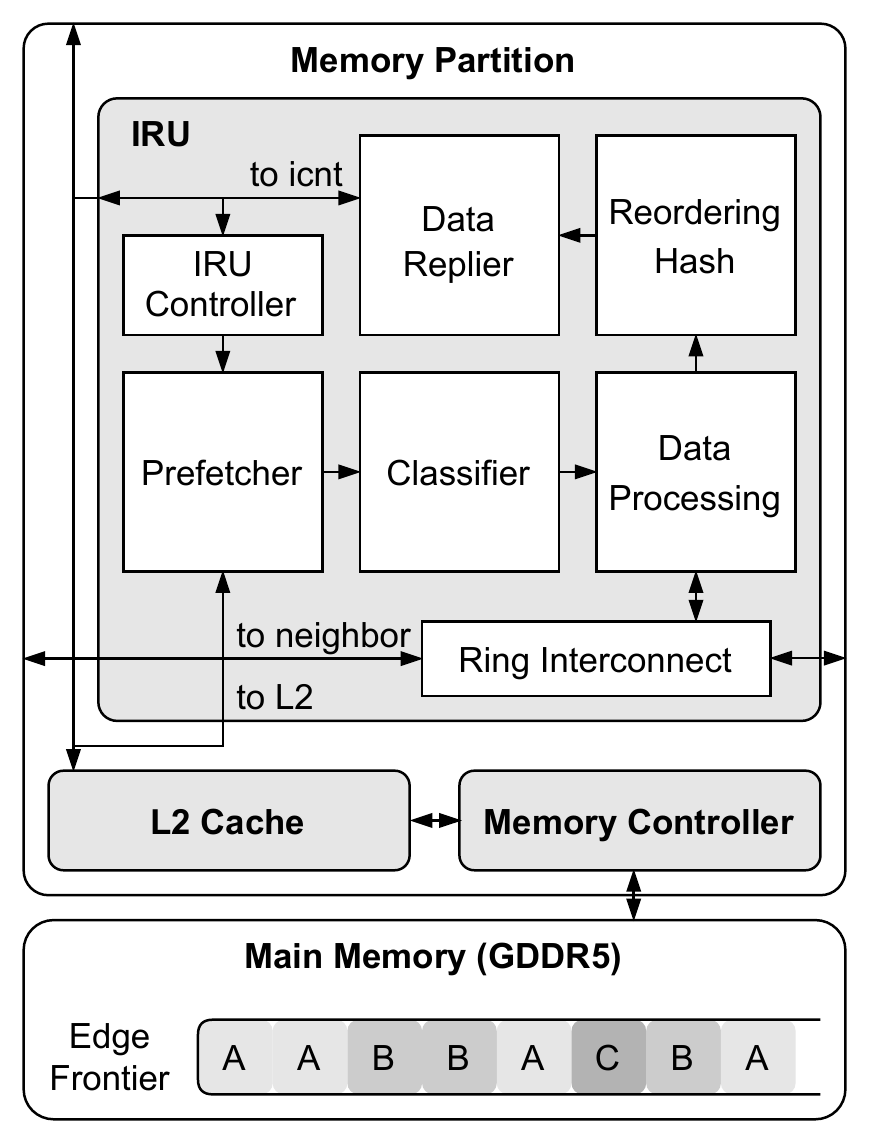}};
        \end{tikzpicture}
        \vspace{-0.15in}
        \subcaption{Overview of the IRU internal pipelined architecture.}
        \label{f:iru_hw}
    \end{subfigure}%
    \hfill
    \begin{subfigure}[h]{0.615\linewidth}
        \begin{subfigure}[t]{0.5\linewidth}
            \centering
            \begin{tikzpicture}
            \node[anchor=south west,inner sep=0] at (0,0) {\includegraphics[width=0.9\linewidth]{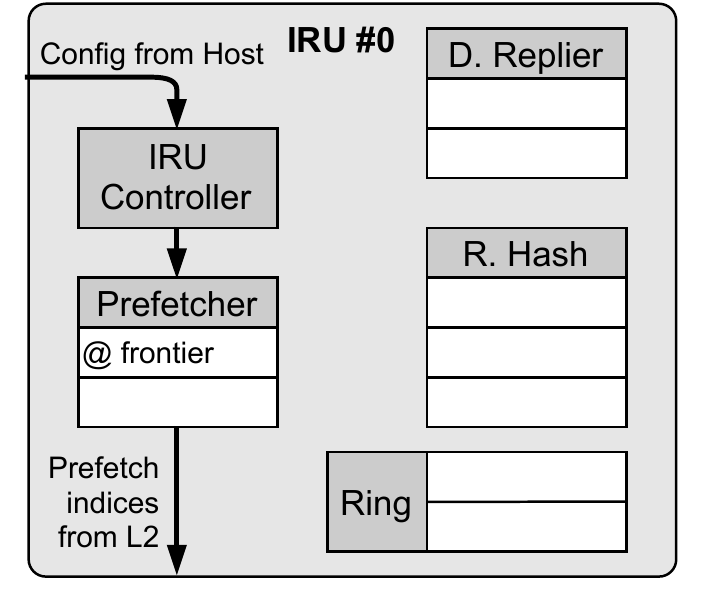}};
            \node at (0.27\linewidth,0.645\linewidth) {\ding{202}};
            \node at (0.27\linewidth,0.15\linewidth) {\ding{203}};
            \end{tikzpicture}
            \vspace{-0.1in}
            \subcaption{Configuration and Prefetch.}
            \label{f:iru_process_config}
        \end{subfigure}%
        \hfill
        \begin{subfigure}[t]{0.5\linewidth}
            \centering
            \begin{tikzpicture}
            \node[anchor=south west,inner sep=0] at (0,0) {\includegraphics[width=0.9\linewidth]{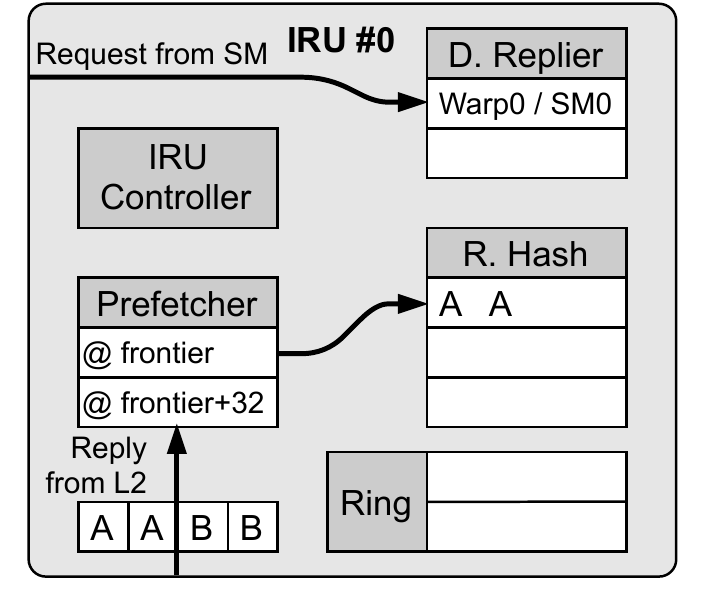}};
            \node at (0.27\linewidth,0.18\linewidth) {\ding{204}};
            \node at (0.5\linewidth,0.35\linewidth) {\ding{205}};
            \node at (0.1\linewidth,0.645\linewidth) {\ding{206}};
            \end{tikzpicture}
            \vspace{-0.1in}
            \subcaption{Data and Request retrieval.}
            \label{f:iru_process_data}
        \end{subfigure}%
        \hfill\null
        \vspace{0.05in}
        \begin{subfigure}[t]{0.5\linewidth}
            \centering
            \begin{tikzpicture}
            \node[anchor=south west,inner sep=0] at (0,0) {\includegraphics[width=0.9\linewidth]{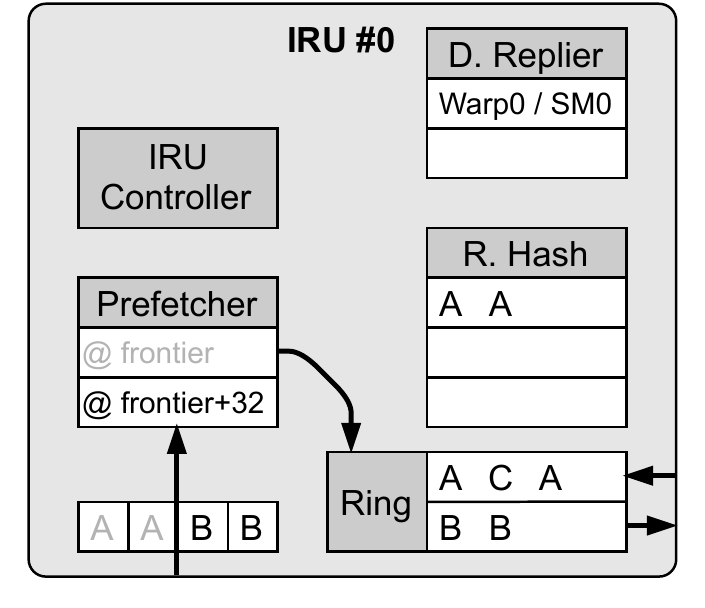}};
            \node at (0.4\linewidth,0.26\linewidth) {\ding{207}};
            \node at (0.76\linewidth,0.135\linewidth) {\ding{208}};
            \end{tikzpicture}
            \vspace{-0.1in}
            \subcaption{Ring interconnection.}
            \label{f:iru_process_ring}
        \end{subfigure}%
        \hfill
        \begin{subfigure}[t]{0.5\linewidth}
            \centering
            \begin{tikzpicture}
            \node[anchor=south west,inner sep=0] at (0,0) {\includegraphics[width=0.9\linewidth]{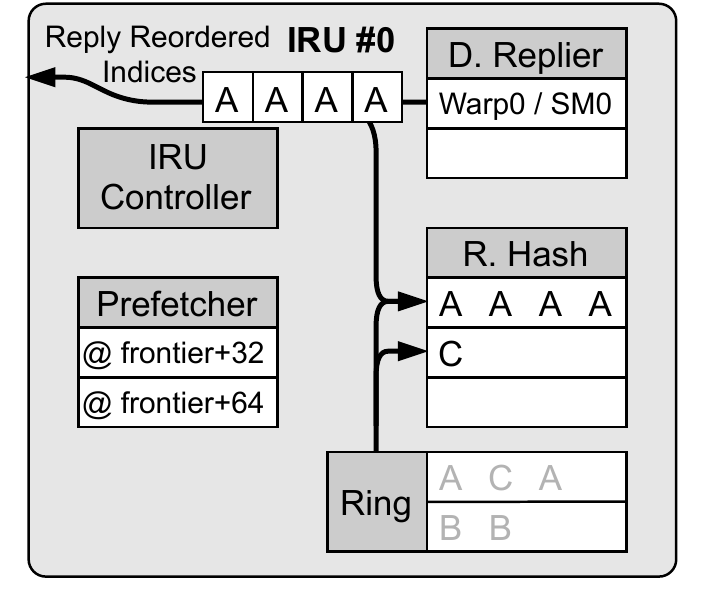}};
            \node at (0.43\linewidth,0.57\linewidth) {\ding{209}};
            \end{tikzpicture}
            \vspace{-0.1in}
            \subcaption{Requests reply.}
            \label{f:iru_process_reply}
        \end{subfigure}%
    \end{subfigure}%
\caption{Architecture and the internal processing performed by the IRU. The indices in memory (from Figure~\ref{f:iru_coalescing_improvement}) are processed by two IRU partitions (IRU 0 shown), which are later replied to a request coming from Warp 0 in SM 0. }
\label{f:iru_process_figure}
\end{figure*}
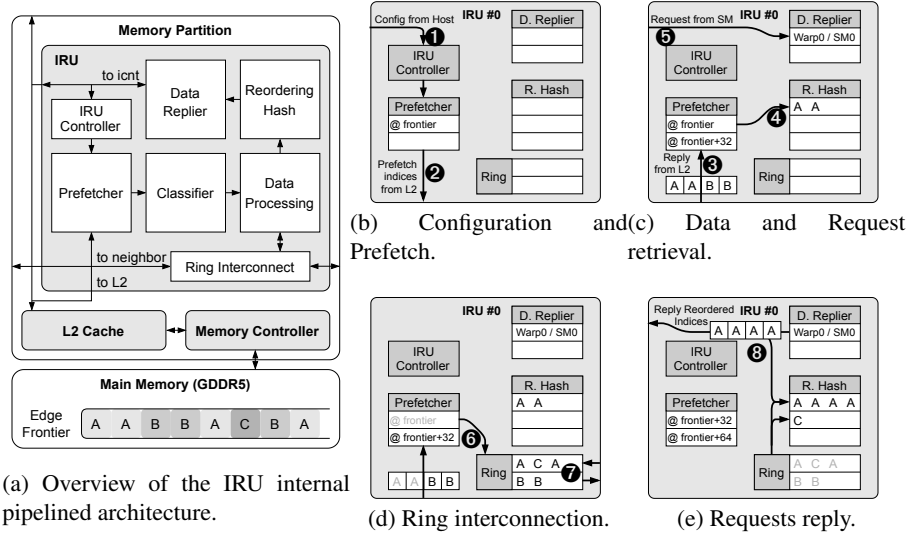

The hardware architecture of the IRU is shown in Figure~\ref{f:iru_hw}. The main purpose of the IRU, which is to reorder indices to improve memory coalescing, is accomplished with the use of a hash table located inside the \textit{Reordering Hash} block. The IRU is integrated inside the memory partitions of the GPGPU architecture, i.e. together with the L2 cache partition, the atomic operations unit and the memory controller. GPUs include multiple memory partitions, in our proposal each partition will contain an instance of the IRU. Instead of having multiple private hash tables, there is a single logical hash table partitioned among the IRUs. This motivates the inclusion of a ring interconnection between the IRUs to forward the data to the corresponding partition of the logical hash table. We have observed that the degree of memory coalescing is significantly affected if each IRU hash table is private and separated; which would constrain IRUs reordering scope to data from a single memory partition. Finally, requests are issued to the L2 to exploit data locality among kernel executions. Alternatively, requests can be configured to bypass L2, which could be beneficial for streaming kernels.




The overall internal processing of the IRU is shown in Figure~{\ref{f:iru_process_figure}}. The figure covers a general overview of the internal IRU architecture and the detailed step by step working of the most relevant components of the IRU covering: configuration and prefetching ({\ref{f:iru_process_config}}), data and requests retrieval ({\ref{f:iru_process_data}}), ring interconnection interaction ({\ref{f:iru_process_ring}}) and requests reply ({\ref{f:iru_process_reply}}).

\subsubsection{Prefetching and Data Processing:}

The \textit{IRU Controller} is initialized by the Host by executing the \textit{configure\_iru} function with the corresponding data~{\ding{202}}. Later the \textit{Prefetcher} uses this data to determine the addresses to prefetch when the GPU kernel starts execution~{\ding{203}}. The \textit{Prefetcher} issues a limited number of on-the-fly prefetch requests to avoid saturating memory bandwidth. Each IRU only prefetches information from its corresponding memory partition. In Figure~{\ref{f:iru_process_figure}}, the first four elements from main memory are fetched by IRU 0, while the next four by IRU 1. When a reply comes back, the retrieved data is stored in a FIFO queue to be later processed.

Afterwards, the \textit{Classifier} block processes the prefetched data~{\ding{204}} by splitting it into smaller FIFO queues, with a throughput of one element per cycle on each queue. The smaller FIFO queues contain the elements that will be inserted in the hash or forwarded through the ring. A hashing function of the element is used to determine which hash table entry it is mapped to and, therefore, if it will access a local bank or must be sent through the ring. Finally, the \textit{Data Processing} block retrieves elements from both the smaller FIFO queues and the ring, prioritizing the latter, and forwards them to the ring or inserts them into the local hash table~{\ding{205}}. On Figure~{\ref{f:iru_process_reply}}, the elements labeled \textit{A} are inserted into the local hash table, as they are determined to access the same memory block.

Meanwhile, requests from the SMs can be received at any time which are then processed by the \textit{Data Replier}~{\ding{206}}. This request originates directly from the SM (i.e. bypassing the L1) and are generated by the extended ISA \textit{load\_iru} operations, that are responsible to retrieve the IRU processed data. Their information is stored until enough data is available to satisfy the request or until a timeout is reached. 

\subsubsection{Ring and Data Reply:}

Due to the partition of the reordering hash table, the hash function of the elements fetched from a memory partition can require that element to be inserted in another IRU partition. The \textit{Ring Interconnection} allows to receive and send elements to the neighbor partitions at every cycle. In Figure~{\ref{f:iru_process_ring}}, the elements labeled \textit{B} are determined to correspond to another IRU partition and so are inserted in the ring~{\ding{207}}. Meanwhile, data from the neighbor partition is received (indices \textit{A} and \textit{C} correspond to IRU 0)~{\ding{208}}.

Lastly, the elements corresponding to this IRU partition are gathered from the ring and inserted into the reordering hash table. When the \textit{Data Replier} detects a hash entry that is complete, or enough data is available to reply a request, the oldest request is replied back to the SMs with that entry's reordered elements~{\ding{209}}, and the data is evicted from the hash table. The data used for the reply (four \textit{A}) are the indices used for the irregular access being optimized. Additionally, more data might be processed per element, requiring at most two replies to be issued, as some algorithms require extra data associated with each index.

Additionally, a timeout is employed to avoid excessively delaying a request. Once the timeout is reached, it then fetches data from the hash table with the best coalesced data entry present, and sends the response once enough data to satisfy the request is retrieved, effectively trading-off worse coalescing for lower latency. Furthermore, simple control logic is added to the SM and IRU partitions to handle balancing issues (i.e between request and entries ready). Each SM distributes the requests evenly across the different IRUs in the memory partitions, and requests can be replied by IRU partitions other than the original. Finally, when no more data is left to be inserted into the IRU, the \textit{Data Replier} replies to the SM by intelligently merging the remaining hash entries. 

\subsection{Reordering Hash}

\begin{figure}[t!]
    \centering
    \includegraphics[width=0.5\linewidth]{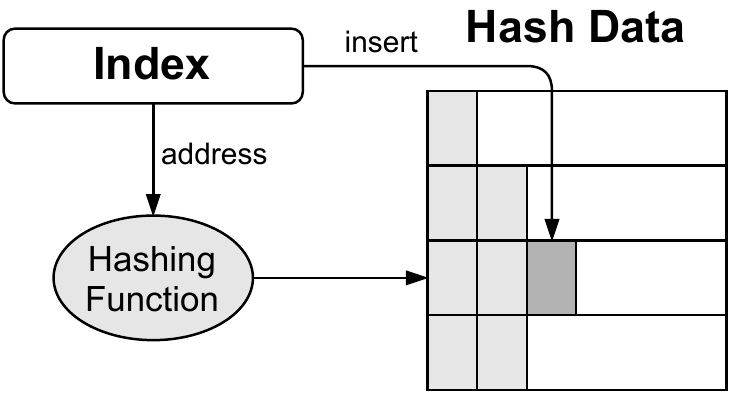}
    \caption{Hash table insertion diagram showcasing how the element is used for the hashing function and is stored in the hash table data.}
    \label{f:iru_hash}
\end{figure}

The \textit{Reordering Hash} contains a physical partition of the global logical hash, which is direct mapped and multi-banked. Each entry holds up to 32 elements that are inserted into the entry in subsequent locations at every hash insertion, as depicted in Figure~\ref{f:iru_hash}.
Furthermore, the hash function key that points to an entry is generated from the value being inserted into the hash table entry. The computation of the hash function collocates in a single hash table entry the elements that will generate memory fetches that target the same memory block, which provides the memory coalescing improvement achieved with the IRU.

Unlike a regular hash table, an insertion allows to merge elements into a hash entry even if the tag does not match. The inherent drawback of this decision is that the elements that a hash table entry collocates might actually not access the same memory block, and thus the memory coalescing that it can achieve will not be optimal. Nonetheless, this design decision largely reduces hardware complexity. Furthermore, a good dispersion hash function and properly sized hash tables limits the amount of conflicts and the effects on memory coalescing. 
Ultimately, when an entry is completely filled with 32 elements, no more data can be inserted to it. At this point, it has 32 collocated elements that potentially will access the same memory block when the program uses them to perform an irregular access, unless there were conflicts. Note that some of these conflicting elements might collocate among themselves, thus not severely impairing memory coalescing.

Some API operations described in Section~\ref{s:api_unit} require additional comparators or adders to be used in a hash table insertion. The additional data that the elements might have is processed by this hardware, which effectively merges or filters an element present in the hash table with the one being inserted. Since these operations will filter out elements, some threads that requested data will not receive any element, which is handled by the \textit{Data Replier} and exposed to the programmer with the API. 

\section{IRU Programmability}\label{s:api_unit}

Ease of programability is a crucial aspect to write efficient parallel programs, a reason for which toolkits such as CUDA are very successful. As described in Section~\ref{s:background}, efficient irregular programs require complex optimization techniques. The IRU has been designed to be easily programmable, so existing codes can exploit this new hardware unit with minimal changes. The IRU extends the GPGPU ISA to support memory load operations that fetch data from the IRU, which require small changes to the pipeline to decode these instructions, in addition to minor changes to the LD/ST unit to route these requests to the IRU. To avoid directly using assembly instructions, we provide a simple API with functions that can be called from CUDA kernels. Furthermore, since the changes to the code are minimal, a compiler that supports the ISA extensions can issue the new instructions when appropriate, freeing the programmer from performing the optimization effort and delivering more code for irregular applications.

IRU's main optimization is the reordering of indices fetched from memory that are later used for irregular accesses. This optimization is based on the premise that the assignment of data to the threads can be safely changed, i.e. each data element (e.g. node/edge in a graph application) can be processed by any thread. Consequently, to be able to correctly utilize the IRU for this optimization, the programmer has to guarantee that the reordering can be applied correctly. The API provides additional functionality used to indicate when it is safe to replace a regular load by an IRU load instruction.

\begin{figure}[t!]
     \centering
     \includegraphics[width=0.45\linewidth]{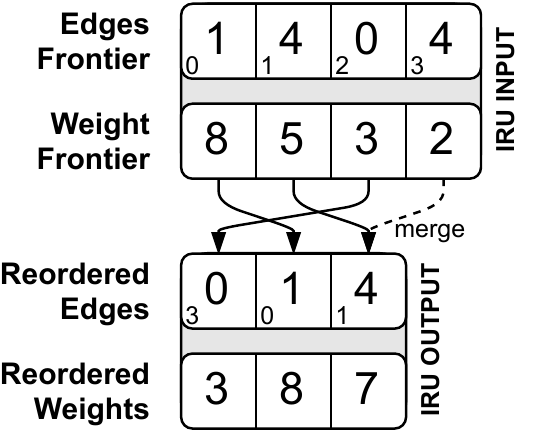}
     \caption{IRU processing of two arrays with filtering enabled. The edges frontier represents the array of indices, while the weight frontier is the secondary array. Filtering is an additional operation of the IRU that can be enabled to remove duplicated elements in order to reduce workload.}
     \label{f:iru_api_data}
     \vspace{-0.2in}
\end{figure}

The baseline functionality provided by the API and IRU hardware supports reordering of an array of 24-bit indices. Additionally, a secondary 32-bit array can be processed simultaneously, yet the reordering is based on the indices array as to improve the coalescing achieved when performing an irregular access. The data (i.e. index and entry in the secondary array) provided to the threads is reordered applying the same reordering to both indices and secondary array, maintaining the original pair of index and secondary data.
Figure~\ref{f:iru_api_data} shows how the input data, first two rows, is reordered in the output data, last two rows. 
The reordering is based on the array of indices, the edge frontier, and every edge is kept with its corresponding weight. This secondary array can be used to process attributes or extra data of the elements being processed. It might be the case that more than a single additional array has to be processed in some application. In this case, the reordering operation can return in which position in the original array the reordered element was located. 
This position can be used to fetch any additional attributes required from multiple arrays.

Graph-based algorithms process many nodes and edges in parallel. Since it is common that several edges lead to the same destination node, many duplicated nodes may appear in the node frontier, producing redundant work in subsequent iterations of the algorithm. This additional work is usually benign as the program implements filtering techniques, which are effective yet computationally costly due to synchronization requirements. To help the programmer remove this additional workload, the IRU is extended to provide filtering or merging of elements (i.e. pair of index and attribute). The IRU can easily detect duplicated indices that are processed simultaneously and so it can remove them or might perform some operation to merge both elements. The operations supported by the IRU are integer comparison and floating point addition. 
Figure~\ref{f:iru_api_data} shows the merging of two indices into one on the output data by adding their attributes in the secondary array.
Filtering out elements causes some threads to not receive any data, and so we extend the API to indicate if a given thread's data has been filtered out. IRU groups the disabled threads in warps rather than distributing them across many warps, this approach allows to minimize branch divergence, remove redundant work and improve performance.

\lstset{style=api_code}
 \begin{figure}[h!]
     \centering
\begin{mdframed}[backgroundcolor=backcolour]
\begin{lstlisting}[language=C++]
void configure_iru (
  addr_t target_array,
  size_t target_array_data_type_size,
  addr_t indices_array, addr_t secondary_array,
  size_t number_elements, filter_op_t filter_op );

__device__ bool load_iru (
  addr_t &indices_array, addr_t &secondary_array,
  uint32_t &position );
\end{lstlisting}
\end{mdframed}
     \caption{IRU API function declarations.}
     \label{f:code_api_description}
\end{figure}

The IRU API, shown in Figure~\ref{f:code_api_description},
provides two main functions: $configure\_iru$, used from the host to configure the IRU, and $load\_iru$, used inside the CUDA kernel to retrieve reordered data from the IRU.
At the start of kernel execution, the $configure\_iru$ function is called to provide all the parameters of the data that will be processed. The required parameters are: target array base address and data type width, both parameters used to configure the offset to be applied to the indices as to compute the coalescing required; the indices array is required too, which is the main data reordered; and finally, the number of elements in the indices array. Optional parameters include the additional secondary array, reordered together with the indices array, and the optional filtering operation performed.
The memory load operation replaces regular load instructions, retrieves the original position of the indices and indicates if a thread is disabled.

\subsection{IRU enabled Graph Applications}
The previously described API enables the instrumentation of state-of-the-art graph-based algorithms such as BFS, SSSP and PR. Although we use push graph implementations, the IRU is not specifically targeting push or pull. The ease of use of our API allows very simple instrumentation an minimal code changes while providing efficient memory coalescing improvements. The following examples show how $load\_iru$ can be used within GPGPU kernels to easily replace existing code.

\lstset{style=cuda_code}
\begin{figure}[h!]
    \centering
\begin{mdframed}[backgroundcolor=backcolour]
\begin{lstlisting}[language=C++,linerange={2-13}]
__global__ void BFS_Contract (...) {
	int pos = blockDim.x * blockIdx.x + threadIdx.x;
	if (pos < number_elements) {
		int edge;
#ifdef NOT_INSTRUMENTED
 		edge = edge_frontier[pos];
#elif USE_IRU
		load_iru(edge);
#endif
		label[edge] = distance;
	}
}
\end{lstlisting}
\end{mdframed}
    \caption{Instrumentation of a BFS Kernel using the IRU.}
    \label{f:code_bfs_iru}
\end{figure}

The basic functionality of the IRU is a good fit for the BFS algorithm as illustrated in Figure~\ref{f:code_bfs_iru}. The indices found in the $edge\_frontier$ array are used to access the $label$ array, resulting in irregular memory accesses and poor memory coalescing. The programmer can easily replace the previous instruction with the $load\_iru$ operation to obtain the indices in such a way that memory coalescing is improved and thus overall performance increases.

\begin{figure}[h!]
    \centering
\begin{mdframed}[backgroundcolor=backcolour]
\begin{lstlisting}[language=C++,linerange={2-16}]
__global__ void SSSP_Compaction (...) {
	int pos = blockDim.x * blockIdx.x + threadIdx.x;
	if (pos < number_elements) {
		int edge, weight;
#ifdef NOT_INSTRUMENTED
		edge = edge_frontier[pos];
		weight = weight_frontier[pos];
#elif USE_IRU
		load_iru(edge, weight, pos);
#endif
		int old = atomicMin(&label[edge], weight);
		if (old > weight)
			lookup[edge] = pos;
	}
}
\end{lstlisting}
\end{mdframed}
    \caption{Instrumentation of an SSSP Kernel using the IRU.}
    \label{f:code_sssp_iru}
\end{figure}

The SSSP algorithm processes additional data per element; each edge has an associated weight value. Figure~\ref{f:code_sssp_iru} shows how $load\_iru$ can handle the use of an additional array, while also retrieving the original position of the reordered element in the $pos$ variable. Note that the algorithm requires the $pos$ variable to be correctly updated with the reordered element in line 17, which is easily accomplished with our API extension.

\begin{figure}[h!]
    \centering
\begin{mdframed}[backgroundcolor=backcolour]
\begin{lstlisting}[language=C++,linerange={2-17}]
__global__ void PR_Contract (...) {
	int pos = blockDim.x * blockIdx.x + threadIdx.x;
	if (pos < number_elements) {
		int edge; float weight;
		bool active_thread = true;
#ifdef NOT_INSTRUMENTED
		edge = edge_frontier[pos];
		weight = weight_frontier[pos];
#elif USE_IRU
		active_thread = load_iru(edge, weight);
#endif
		if (active_thread)
			atomicAdd(&label[edge], weight);
	}
}
\end{lstlisting}
\end{mdframed}
    \caption{Instrumentation of a PageRank Kernel using the IRU.}
    \label{f:code_pr_iru}
\end{figure}

Finally, the PageRank kernel shown in Figure~\ref{f:code_pr_iru} performs additions of the elements' weights into the $label$ array. Utilizing the filtering/merge functionality of the IRU, an initial addition can be performed while the elements are being processed in the IRU, which allows to disable merged out threads. The $load\_iru$ function returns whether or not the thread has a valid element or if it has been merged out; the value in a retrieved element's $weight$ has the sum of those $weight$ of the same $edge$. Note that the filtering is not complete as it merges only elements found concurrently on the IRU, yet it manages to filter a significant amount of duplicated elements. Overall, this extension allows reducing the workload of the kernel, in this case, reducing the number of $atomicAdd$ required.

\section{Evaluation Methodology}\label{s:methodology}

We have implemented the IRU architecture in GPGPU-Sim~\cite{bakhoda2009analyzing}. We extended the memory partitions in GPGPU-Sim to accurately model IRU's hardware as shown in Figure~\ref{f:iru_hw}. Furthermore, we extended the Streaming Multiprocessors (SMs) to support our new instructions.

Each partition of the IRU uses a 2 KB FIFO to buffer warp requests and 1.7 KB for the prefetching buffer (8 on-the-fly prefetches). A buffer of 1.2 KB is used for the Classifier block to determine the data destination. The ring requires a total of 2.8 KB buffering. The main component of the IRU is the direct-mapped hash table with 1024 sets, split in 4 physical partitions. Each partition is 2-way banked, holding 256 sets, requiring 80 KB of total storage, significantly smaller than the 512 KB of the L2 partition. Table~\ref{t:iru_hw} summarizes the components of an IRU partition. Since the IRU is mostly comprised of SRAM elements without complex execution units we model area and energy using CACTI~\cite{li2009mcpat} with a node technology of 32 nm.

\begin{table}[ht!]
\small
\centering
\caption{IRU hardware requirements per partition.}
\label{t:iru_hw}
\begin{tabular}{rl}
    \hline
    
    \textbf{Component}      & \textbf{Requirements} \\
    
    \hline
    
    Requests Buffer         & 2 KB          \\
    Prefetcher Buffer       & 1.7 KB        \\
    Classifier Buffer       & 1.2 KB        \\
    Ring Buffer             & 2.8 KB        \\
    Hash Data        & 80 KB                \\
    
    \hline
\end{tabular}
\end{table}

GPGPU performance is modeled with GPGPU-Sim~\cite{bakhoda2009analyzing}, energy consumption and area with GPUWattch~\cite{leng2013gpuwattch}, both simulators configured with the parameters shown in Table~\ref{t:gtx980} to model an NVIDIA GTX 980. To evaluate our proposal we use state-of-the-art GPGPU implementations of BFS~\cite{merrill2015high}, SSSP~\cite{davidson2014work}, and PageRank~\cite{geil2014wtf} graph algorithms. We run these graph processing algorithms with datasets representative of different application domains with varied sizes, characteristics and degrees of connectivity, shown in Table~\ref{t:datasets}, and collected from well-known repositories of research graph datasets~\cite{davis2011university, dimacs2010}.

\begin{table}[ht!]
\small
\centering
\caption{Parameters employed in the experiments to model a GTX 980 in GPGPU-Sim.}
\label{t:gtx980}
\begin{tabular}{rl}
    \hline
    
    \textbf{Characteristic}                       & \textbf{Configuration}     \\
    
    \hline
    
    GPU, Frequency                       & NVIDIA GTX 980, 1.27GHz     \\
    Streaming Multiproc. & 16 (2048 threads), Maxwell        \\
    L1, L2 caches             & 32 KB, 2 MB. 128 B lines          \\
    Memory Partitions        & 4 (4 channel GDDR5)                 \\
    Main Memory               & 4 GB GDDR5, 224 GB/s    \\
    
    \hline
\end{tabular}
\end{table}

\begin{table*}[th!]
\small
\centering
\caption{Benchmark graph datasets.}
\label{t:datasets}
\begin{tabular}{llccc}
    \hline
    
    \textbf{Graph Name} & \textbf{Description} & \textbf{Nodes (10$^3$)} & \textbf{Edges (10$^6$)} & \textbf{Avg. Degree} \\
    
    \hline
    ca~\cite{davis2011university}       & California road network          & 710    & 3.48    & 9.8   \\
    cond~\cite{davis2011university}     & Collaboration network, arxiv.org & 40     & 0.35    & 17.4  \\
    delaunay~\cite{dimacs2010}          & Delaunay triangulation           & 524    & 3.4     & 12    \\
    human~\cite{davis2011university}    & Human gene regulatory network    & 22     & 24.6    & 2214  \\
    kron~\cite{dimacs2010}              & Graph500, Synthetic Graph        & 262    & 21      & 156   \\
    msdoor~\cite{davis2011university}   & Mesh of 3D object                & 415    & 20.2    & 97.3  \\
    
    \hline
\end{tabular}
\end{table*}

\section{Experimental Results}\label{s:results}

In this section, we evaluate the performance and energy efficiency of our IRU hardware presented in Section~\ref{s:hw}. More specifically, we analyze how the memory hierarchy contention is reduced, the reduction of interconnection traffic, the improvement on memory coalescing, the IRU filtering capabilities, and the overall speedup and energy savings of our proposed GPU+IRU with respect to the baseline GPU.

\subsection{Memory Pressure Reduction}

\begin{figure*}[t!]
\centering
\includegraphics[width=1\linewidth]{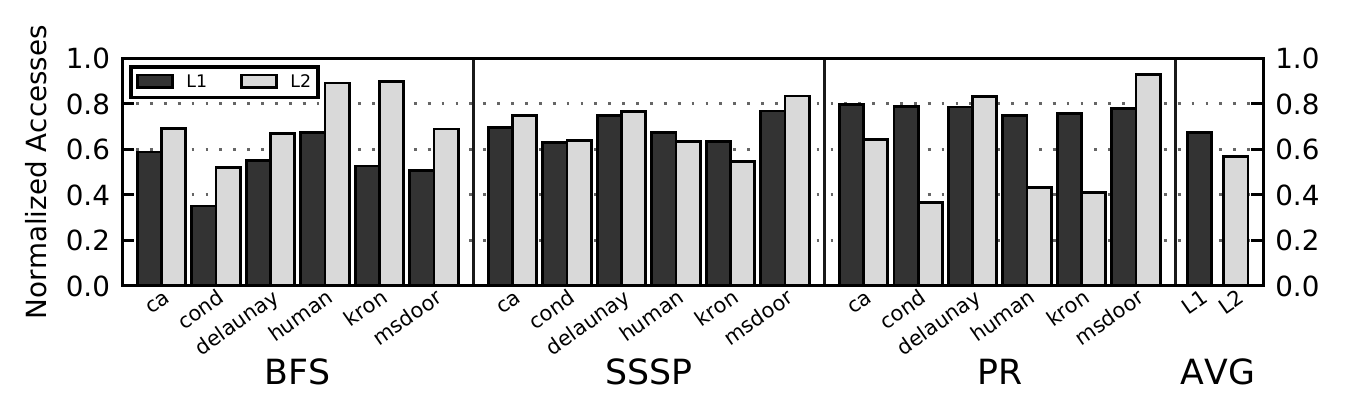}
\null
\caption{Normalized accesses to L1 and L2 caches of the IRU enabled GPU system versus the Baseline GPU system (GTX 980 GPU with parameters shown in Table~\ref{t:gtx980}). Significant reductions are achieved across BFS, SSSP and PR graph algorithms and every dataset.}
\label{f:results_l1l2_access_reduction}
\end{figure*}

IRU's main functionality is to reorder irregular accesses improving their memory coalescing thus reducing the overall contention in the memory hierarchy. Figure~\ref{f:results_l1l2_access_reduction} shows how the IRU consistently reduces accesses and contention on both L1 and L2 across all graph algorithms and datasets. Accesses to L1 and L2 are reduced to as low as 35\% and 36\% for the \textit{cond} benchmark on BFS and PR respectively. Overall, accesses are reduced to 67\% and 56\% for L1 and L2 caches on average. 

This important reduction comes from several factors. First, the IRU reordering of irregular accesses improves coalescing reducing the accesses to L1. Second, IRU reorders requests across SMs so it collocates accesses of a particular memory block to a single SM, avoiding data replication across L1 data caches, improving hit ratios. Third, reduced accesses to L1 avoid capacity and conflict misses, improve data thrashing and consequently reduce L2 accesses. Finally, IRU filtering further reduces accesses by removing/merging duplicated elements, that avoids additional memory accesses. 

L2 accesses reduction is greater than in L1 in some benchmarks for SSSP and PR graph algorithms. Many indices reordered by the IRU on SSSP and PR are used for irregular accesses performed by atomic instructions. In GPGPU-Sim atomic operations bypass the L1 and are handled at the memory partitions. IRU coalescing and filtering improvement for these operations reduces L2 accesses but not L1 accesses, explaining the larger reduction in L2 accesses compared to L1 for SSSP and PR. Note that atomic operations within a warp are coalesced as long as different threads access different parts of a cache line.

\begin{figure*}[t!]
\centering
\includegraphics[width=1\linewidth]{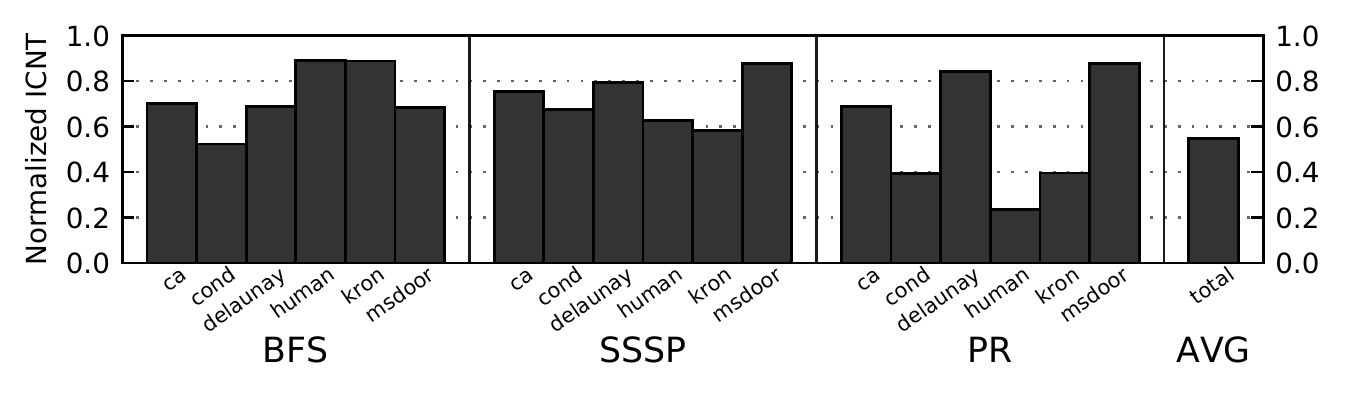}
\null
\caption{Normalized interconnection traffic between SM (Streaming Multiprocessors) and MP (Memory Partitions) for the IRU enabled GPU system over the Baseline GPU system (GTX 980 GPU with parameters shown in Table~\ref{t:gtx980}). Significant reductions are achieved across BFS, SSSP and PR graph algorithms and every dataset.}
\label{f:results_icnt_traffic_reduction}
\end{figure*}

We have also analyzed the impact of the IRU in the Network-on-Chip (NoC) that interconnects the Streaming Multiprocessors (SM) with the Memory Partitions (MP). Figure~\ref{f:results_icnt_traffic_reduction} shows the normalized traffic in the NoC. As it can be seen, the IRU consistently reduces interconnection traffic across all graph algorithms and datasets. Traffic between SM and MP is reduced to as low as 23\% for the \textit{human} benchmark on PR, overall reducing NoC traffic to 54\% of the original interconnection traffic. This reduction is due to several factors. First, the improved memory coalescing results in a more efficient use of the L1 data cache, significantly reducing the number of misses. Second, filtering also contributes to lower L2 accesses which reduces interconnection contention. Finally, the extended ISA instructions allow reduced traffic by issuing a single request to the IRU that receives two replies, whereas the baseline GPU would have issued two requests and two replies in order to gather data in different frontiers. Note that the IRU API allows to gather an index together with an extra attribute (e.g. weight of edge in a graph) with just one memory request as explained in Section~\ref{s:api_unit}.

\begin{figure}[t!]
\centering
\includegraphics[width=0.6\linewidth]{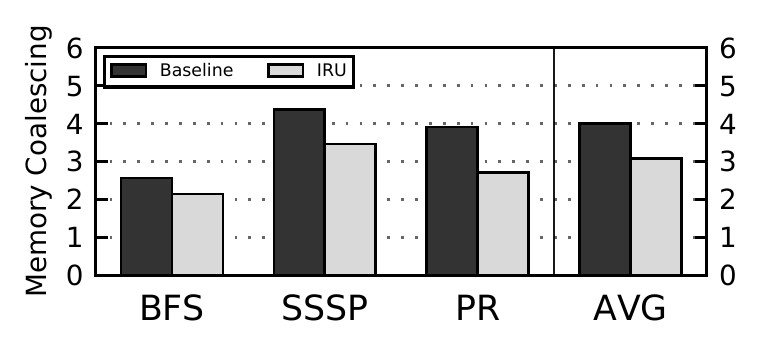}
\caption{Improvement in memory coalescing achieved with the IRU over the Baseline GPU system (GTX 980 GPU with parameters shown in Table~\ref{t:gtx980}). Vertical axis shows the number of memory requests sent to the L1 cache on average per each memory instruction.}
\label{f:results_coalescing_improvement_summary}
\end{figure}

Figure~\ref{f:results_coalescing_improvement_summary} shows the improvement in memory coalescing delivered by the IRU. A higher number indicates that more accesses are required to serve each warp memory request, with a maximum of 32 accesses per request, and a minimum of 1 access in the best scenario. The IRU improves the overall coalescing for every graph algorithm from 4 to 3 accesses per memory requests on average. This improvement is significant given that the filtering schemes that some of the algorithms employ, combined with the filtering applied by the IRU, reduce the potential for coalescing memory requests, since filtering removes some duplicated elements whose accesses could be coalesced. Nonetheless, memory coalescing is significantly improved, reducing the pressure on the memory hierarchy.

Finally, main memory accesses are reduced by 4\% due to reduced L2 misses as a result of reduced accesses. Overall, reordering and filtering techniques allow the IRU to deliver significant improvements in memory coalescing and reduce contention in multiple levels of the memory hierarchy.

\subsection{Filtering Effectiveness}

\begin{figure}[t!]
 \centering
 \includegraphics[width=0.65\linewidth]{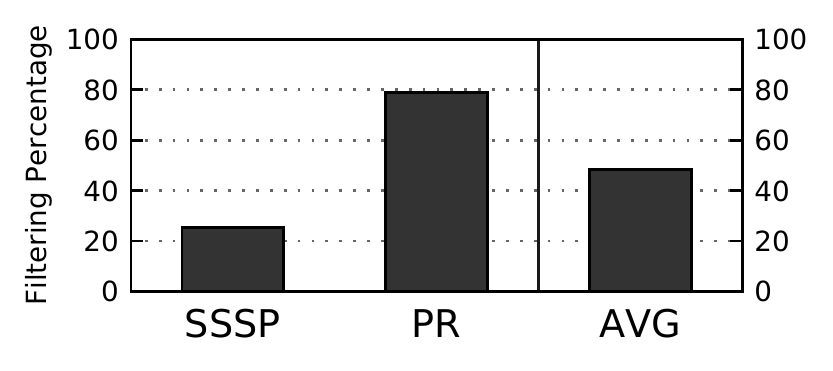}
 \caption{Percentage of elements that are filtered out in our IRU-enabled GPU system.}
 \label{f:results_filtering_percentage_summary}
 \vspace{-0.05in}
\end{figure}

The IRU hardware provides filtering capabilities without complex additional hardware. Figure~\ref{f:results_filtering_percentage_summary} shows the percentage of elements (i.e. indices with their adjacent data) processed by the IRU which are filtered out or merged. 
We apply the filtering to both SSSP and PR, achieving 23\% and 79\% workload filtering respectively. On average, 48.5\% of the elements are filtered out by the IRU.
Note that this high percentage does not directly indicate that a similar amount of accesses to memory are avoided with respect to the baseline GPU, as state-of-the-art CUDA implementations of SSSP and PR include sophisticated mechanisms to filter our duplicated elements. However, in our proposed scheme the filtering is performed by the IRU hardware, whereas in the baseline GPU this filtering process is done in software. Hence, our proposal is effective at filtering/merging duplicated elements by leveraging the IRU hardware that is already available for the reordering operation, avoiding costly software filtering schemes of graph algorithms.


\subsection{Performance and Energy Evaluation}

\begin{figure*}[t!]
\centering
\includegraphics[width=1\linewidth]{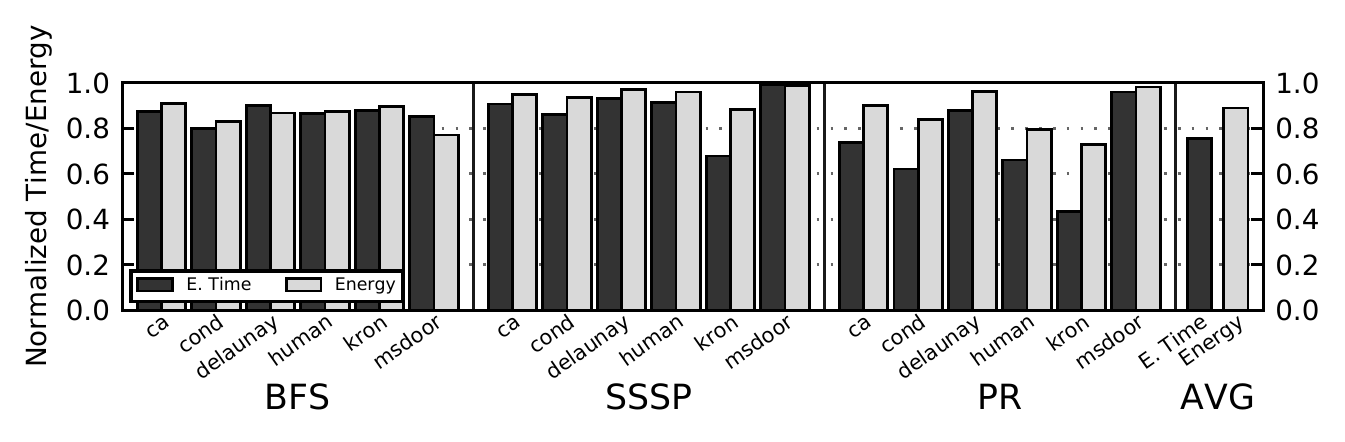}
\null
\caption{Normalized execution time and energy consumption reduction of the IRU enabled GPU with respect to the baseline GPU system (GTX 980 GPU with parameters shown in Table~\ref{t:gtx980}). Significant speedups and energy savings achieved across BFS, SSSP and PR graph algorithms and every dataset.}
\label{f:results_time_energy}
\end{figure*}

The IRU provides performance improvements across all algorithms and benchmarks, as shown in Figure~\ref{f:results_time_energy}. On average the IRU achieves a speedup of 1.33x, with average speedups of 1.16x, 1.14x and 1.40x for BFS, SSSP and PR respectively. PR exhibits higher speedups due to larger reduction of L2 accesses achieved by the filtering, which avoids costly atomic L2 accesses. SSSP achieves the lowest speedup due to lower filtering effectiveness.
Overall, performance improvements come from two sources. First, the IRU improved memory coalescing by reordering of indices used for irregular accesses, which reduces contention on the memory hierarchy. Second, the IRU filtering and merging that enables further reduction of memory accesses and avoids wasted cycles in the functional units of the GPU due to processing redundant elements.

Figure~\ref{f:results_time_energy} also shows the energy savings achieved by the IRU, which are significant across all graphs and datasets. On average, the IRU achieves an energy reduction of 13\%, with reductions of 17\%, 5\% and 15\% for BFS, SSSP and PR respectively. Energy savings are more limited than performance improvements since the IRU greatly reduces L1 and L2 accesses but achieves a more modest reduction of main memory accesses. Note that main memory represents a very significant portion of the total energy consumed. The IRU energy overhead represents a small 0.5\% of the final energy.
Overall, energy savings are obtained from several sources. First, the reduced accesses to L1 and L2 and reduced contention in the memory hierarchy. Second, the reduced execution time cuts down on the static power and thus, the overall energy consumption of the GPU system. Third, the energy-efficient IRU which enables the reduction in accesses and contention, and allows more efficient hardware-based filtering than the costly software-based mechanisms employed by graph applications. Finally, the IRU reordering leads to a reduction in main memory accesses which further reduces energy.

\subsection{Area Evaluation}

Our evaluation of the IRU energy and area estimations indicate that the IRU requires a total of 23.9 mm\textsuperscript{2} when adding up all the 4 partitions of our GPU system with a GTX 980, each partition being 5.98 mm\textsuperscript{2}. The entire IRU represents 5.6\% of the total GPU area. Overall, the IRU is a very compact and efficient hardware which manages to deliver significant performance and energy savings with small area requirements.






\section{Related Work}\label{s:related_work}

Irregular programs on GPGPU architectures face many challenges resulting in low GPU utilization and poor performance. Several previous works have thoroughly analyzed the causes of these inefficiencies, that boil down to control flow divergence and memory accesses irregularity~\cite{burtscher2012quantitative,o2014microarchitectural,lumsdaine2007challenges,xu2014graph}. 
Nonetheless, if these issues are overcome, irregular applications can greatly benefit of the high parallelism that GPU architectures offer.
Over the recent years several works have approached the topic of efficient and improved irregular programs on GPGPU architectures.

Some solutions approach the branch divergence issue by providing load balancing solutions~\cite{khorasani2015scalable,merrill2015high} to improve utilization of execution units. Others provide thread remapping over warps to improve branch divergence~\cite{fung2007dynamic}. Some memory divergence approaches propose modifying software data structures~\cite{gharaibeh2013efficient,nodehi2018tigr,wang2016gunrock}, whereas others such as HALO~\cite{gera2020traversing} provide static reordering of a graph to improve data locality.
Many specialized works have focused on GPU execution of irregular Sparse Matrix Vector Multiplication (SpMV) and Matrix Matrix Multiplication (GEMM) by proposing software approaches that reorder the matrices dataset~\cite{pichel2012optimization}, and algorithms tailored for specific matrix data characteristics~\cite{rivera2020ism2}, and row reordering techniques~\cite{jiang2020novel} to improve data locality among processed rows.
Finally, other works prove the NP-completeness of finding the data layouts through data repositioning that minimize uncoalesced memory accesses and propose software algorithms to attain them~\cite{wu2013complexity}.
These works propose methods requiring significant programming effort, as they require changing algorithms and data structures or profound hardware knowledge.
In contrast, our IRU solution requires very lightweight changes of the algorithms and does not require profound knowledge of the inner working of the GPU memory hierarchy to improve memory coalescing and resolve contention issues.

Other approaches explore microarchitectural improvements transparent to the programmer, or with some minor involvement to achieve the desired result. Extensive research has been done on flexible cache solutions~\cite{kumar2012amoeba,li2017elastic,guo2018dycache} that adapt for fine-grained and coarse-grained accesses while other works resort to cache bypassing mechanisms~\cite{chen2014adaptive}.
Works such as LAMAR~\cite{rhu2013locality} explore sizable GPU architecture and memory hierarchy modifications to detect and provide fine and coarse grained accesses throughout the memory system.
Other works propose hybrid software and hardware approaches that enable data dependent aware dynamic scheduling~\cite{yan2019alleviating} or provide prefetching of irregular accesses~\cite{lakshminarayana2014spare} to registers to avoid early data eviction.
Finally, works such as D2MA~\cite{jamshidi2014d2ma} and Stash~\cite{komuravelli2015stash} set to provide mechanisms to manage global data allocation to shared memory, with the objective to increase capacity close to the cores and improve memory hierarchy and overall performance. 
The aforementioned works leverage hardware solutions that work around or ameliorate the consequences of low memory coalescing by providing mechanisms to lower memory contention. In contrast, our IRU provides tools to amend the cause, not the consequence, of the high memory contention, i.e. the poor memory coalescing.
Intermediate approaches have explored extending the GPU architecture with custom purpose hardware units. SCU~\cite{segura2019scu} proposes a programmable GPU hardware extension for graph processing that is tailored to stream compaction operations required for graph processing. 
Meanwhile, the GPU is employed to execute the graph processing workload part that is most well suited for, achieving significant performance improvements.
In comparison, the IRU is a more flexible extension, with a more generic and reusable API tailored to general irregular accesses patterns. Furthermore, the SCU requires significant changes in the application, since entire kernels are replaced by calls to the SCU whereas other kernels must be adapted.
Our solution requires minor changes to the application as described in Section~\ref{s:api_unit}. 

Finally, many works propose to replace entirely the GPU with special purpose accelerators custom-made for graph processing, which set aside the GPU due to fundamental limitations and exploit deep knowledge of graphs data structures. Proposals include standalone approaches such as TuNao~\cite{zhou2017tunao}, Dram-based  Graphicionado~\cite{ham2016graphicionado} or PIM-based GraphH~\cite{dai2018graphh}. In contrast, our IRU solution leverages the popularity of GPU architectures and provides generic solutions that bring the performance and efficiency of GPU architectures for low performing irregular programs.

\section{Conclusions}\label{s:conclusions}

In this paper we propose the Irregular accesses Reorder Unit (IRU), a GPU extension that improves performance and energy efficiency of irregular applications.
Efficient execution of irregular applications on GPU architectures is challenging due to low utilization and poor memory coalescing, which force programmers to carry out complex code optimization techniques to achieve high performance.
The IRU is a novel hardware unit that delivers improved performance of irregular applications by reordering data serviced to threads. This reordering is enabled by relaxing the strict relationship between threads and data processed.
We further extend the IRU to filter out and merge repeated elements while performing the reordering, this results in increased performance by largely reducing redundant GPU workload.
The IRU reordering and filtering schemes deliver 1.32x improved memory coalescing, while reducing the traffic in the memory hierarchy by 46\%. 
Our IRU augmented GPU system achieves on average 1.33x speedup and 13\% energy savings for a diverse set of graph-based applications and datasets, while incurring in a small 5.6\% area overhead.

\section*{Acknowledgment}

This work has been supported by the CoCoUnit ERC Advanced Grant of the EU’s Horizon 2020 program (grant No 833057), the Spanish State Research Agency (MCIN/AEI) under grant PID2020-113172RB-I00 and the ICREA Academia program.

\bibliography{refs}

\end{document}